\documentclass[prd,aps,amsmath,amssymb,letter]{revtex4}
\usepackage{graphicx}% Include figure files
\usepackage{epsfig,rotate,latexsym}
\begin{document}
\title{Using CMBR analysis tools for flow anisotropies 
in relativistic heavy-ion collisions}
\author{Ananta P. Mishra}
\email {apmishra@iopb.res.in}
\author{Ranjita K. Mohapatra}
\email {ranjita@iopb.res.in}
\author{P. S. Saumia}
\email {saumia@iopb.res.in}
\author{Ajit M. Srivastava}
\email{ajit@iopb.res.in}
\affiliation{Institute of Physics, Sachivalaya Marg, 
Bhubaneswar 751005, India}
%
% IOP Preprint number: IP/BBSR/2007-10, Oct 18, 2007. 
%

\begin{abstract}
 Recently we have shown that there are crucial similarities in
the physics of cosmic microwave background radiation (CMBR) anisotropies 
and the flow anisotropies in relativistic heavy-ion collision
experiments (RHICE). We also argued that, following CMBR 
anisotropy analysis, a plot of root-mean square values of the 
flow coefficients, calculated in a lab fixed frame
for RHICE, can yield important information about the nature of initial 
state anisotropies and their evolution. Here we demonstrate the strength of 
this technique by showing that elliptic flow for non-central collisions 
can be directly determined from such a plot without any need for the
determination of event-plane. 
\end{abstract}
\maketitle
%\pacs{PACS numbers: 25.75.-q, 12.38.Mh, 98.80.Cq}
%\keywords: {.....}
%%%%%%%%%%%%%%%%%%%

\section{Introduction}

 Relativistic heavy-ion collision experiments (RHICE) are often
termed as {\it Little Bangs} in analogy of the {\it Big Bang}
representing the initial stage of the Universe. Indeed, there are
tempting similarities between the early universe and these experiments.
For example, it is often mentioned that the surface of last scattering for 
the cosmic microwave background radiation (CMBR) is similar to the freezeout 
surface in RHICE in the sense that one has to learn about the early stages
of the system from hadrons coming out from the freezeout surface, just
as CMBR encodes information about the early universe. However, these
correspondences and analogies have been primarily made from a motivational
view point. It has recently been shown by us \cite{cmbhic} that
the physics of RHICE may have deeper connections, not just with the above 
mentioned aspects of CMBR, but even to the most celebrated aspect of
inflationary physics of the universe namely the presence of superhorizon
fluctuations.

  Following the techniques used for CMBR  anisotropy analysis, it was
proposed in ref.\cite{cmbhic} that instead of focusing on the average
values of the flow coefficients $v_n$, one should calculate root-mean
square values of the flow coefficients $v_n^{rms}$. Further, these
calculations should be performed in a lab fixed frame, which eliminates
the difficulties associated with determination of event plane for
conventional elliptic flow analysis for non-central collisions.
To distinguish from the conventional flow coefficients $v_n$ which
are defined with respect to the event plane, we will denote, in this 
paper, the flow coefficients defined with respect to the lab fixed 
frame as ${\tilde v}_n$. The root mean square value is then denoted
as ${\tilde v}_n^{rms}$.  Plots of ${\tilde v}_n^{rms}$ (denoted
as $v_n^{rms}$ in \cite{cmbhic}) using model estimates based on initial 
energy density anisotropies from HIJING \cite{hijing} were given 
there for large range of values of $n$ (with $n$ ranging from 
1 to 30) for central collisions and it was argued
that such plots can yield important information about the nature of
initial state anisotropies and their evolution for RHICE.

  In this paper we demonstrate the strength of this analysis technique 
by showing that a plot of values of ${\tilde v}_n^{rms}$ vs. $n$ can be 
used for directly probing various flow coefficients, in particular,  
the elliptic flow for non-central collisions \cite{flow0}, without any 
need for the determination of event-plane. For non-central collisions, 
a plot of ${\tilde v}_n^{rms}$  (using model estimates from the spatial
distribution of initial parton energy density), shows a 
prominent peak at $n = 2$, as we will see below. Further, the height 
of the peak increases with increasing value of the impact parameter $b$,
thus allowing direct determination of elliptic flow 
(in terms of ${\tilde v}_2^{rms}$). 
This can be very useful for analyzing huge wealth of data as the 
determination of event plane for each event becomes redundant. 

 The correspondence between the CMBR physics and RHICE explored at
a deeper level as discussed in \cite{cmbhic} may lead to new set 
of interesting phenomena and new techniques for RHICE. We mention 
that such a connection between physics of RHICE and that of inflationary
universe was never anticipated earlier, and indeed, at first sight, 
it looks surprising that a concept like superhorizon fluctuation which 
arises from highly non-trivial, superluminal expansion phase of the very 
early universe could have any relevance for relativistic heavy-ion 
collision experiments in laboratory.
Such superhorizon fluctuations in 
RHICE originate from the fact that in the center of mass 
frame the thermalization happens rather quickly, within about 1 fm. This
time is too short for any fluctuations (inhomogeneities) in the transverse
direction to disappear, by processes of homogenization, 
which have wavelengths larger than 1 fm. Initial parton 
energy density distribution from HIJING  show that 
transverse fluctuations with wavelengths 
significantly larger than 1 fm are necessarily present at the time 
1 fm even in central collisions. These arise from localization of 
partons inside initial nucleons, as well as from the fluctuations in 
nucleon coordinates.  Analyzing the development of flow from such initial
large wavelength fluctuations, we had argued that the anisotropies in the
final particle momenta may show characteristic features of acoustic
oscillations of subhorizon modes as well as suppression of modes which 
remain superhorizon at the freezeout stage. We will show below
that such features do not mask the peak at $n = 2$ in non-central
collisions.

  The paper is organized in the following manner. In section II, we
recall the adoption of CMBR anisotropy analysis method for the case
of RHICE. Here we also show that calculations of ${\tilde v}_n^{rms}$ in 
the lab fixed frame can directly yield information about various flow 
coefficients defined with respect to the event plane. In particular, the
conventional elliptic flow can be determined in this manner. Section III 
presents the results of the application of these analysis methods for 
the determination of ${\tilde v}_n^{rms}$ for the case of non-central 
collisions. In section IV
we discuss the $p_T$ dependence, and in section V we discuss the
issue of anisotropic detector acceptance. Section VI presents 
conclusions and discussions.

\section{Using CMBR analysis tools for RHICE}

Traditional analysis of elliptic flow in RHICE \cite{flow0} is based on 
measuring the average value of the 2nd Fourier coefficient of particle 
transverse momentum anisotropy. The average value 
is determined by carrying out the 
Fourier expansion in the event-plane reference frame. Even for the so-called 
'central collisions' one determines event plane for very small impact
parameter collisions and then carries out the traditional analysis of
determination of average flow \cite{cntrl}.  However, determination of 
event plane is highly non-trivial making determination of higher flow 
coefficients very difficult. Indeed, flow
coefficients have been determined only upto $v_6$. 

  It is here that the use of techniques of CMBR analysis can make a 
significant difference. For CMBR, the 
temperature anisotropies are analyzed using spherical harmonics,
as appropriate for the surface of 2-sphere (the CMBR sky) \cite{cmbr}. 
The coefficients of the expansion (denoted as $a_{lm}$ corresponding to
the spherical harmonic $Y_{lm}$) are degenerate in the argument $m$
and when averaged over different values of $m$ yield zero averages
due to isotropy of the universe (when suitably corrected for
local velocities etc.). What one plots is the variance of $a_{lm}$
denoted by $C_l$ and this leads to the celebrated power spectrum of
CMBR anisotropies \cite{cmbr}.

  We propose using the same technique \cite{cmbhic} for analyzing particle
momentum anisotropies, using lab fixed frame, in RHICE to probe the 
generation and evolution
of flow. For RHICE, focusing on central rapidity region, one will
be analyzing momentum anisotropies on a circle, requiring use of the 
Fourier coefficients, which we denote as ${\tilde v}_n$ to distinguish
from the conventional flow coefficients $v_n$ which are defined with 
respect to the  event plane. With a fixed lab frame, the event 
average values of these ${\tilde v}_n$s will all be zero
due to rotational symmetry. We then propose to use the variance of
${\tilde v}_n$, i.e. ${\tilde v}_n^{rms}$ in analogy with $C_l$ for CMBR. 

 We first establish the relation between ${\tilde v}_n^{rms}$ calculated 
in the lab fixed frame and the conventional flow coefficients $v_n$ which 
are defined with respect to the event plane as follows

\begin{equation}
\rho(\phi^\prime) = \sum_{n=-\infty}^\infty v_n e^{in\phi^\prime}
\end{equation}

 Where the angle $\phi^\prime$ is measured with respect to the event plane.
Assuming that the event plane is oriented by angle $\psi$ with respect to
the lab fixed frame, we can write the above equation in the lab fixed 
frame as,

\begin{equation}
\rho_\psi(\phi) = \sum_{n=-\infty}^\infty v_n e^{in(\phi-\psi)}
\end{equation}
 
 where now the angle $\phi$ is measured with respect to the lab fixed frame.
The subscript for $\rho_\psi(\phi)$ remind us that the above Fourier
series expansion uses the flow coefficients defined with respect to the
event plane which is oriented at an angle $\psi$ in the lab frame. Without the 
knowledge of $\psi$, we will calculate 
the lab fixed flow coefficients ${\tilde v}_n(\psi)$ as

\begin{equation}
 {\tilde v}_n(\psi) =  \frac{1}{2\pi} \int_{-\pi}^{\pi}
\rho_\psi(\phi) e^{-in\phi} d\phi
\end{equation}

 Here the argument $\psi$ in ${\tilde v}_n(\psi)$ is used to remind that 
these quantities are calculated in the lab fixed frame, and are different 
from the values of conventional flow coefficients $v_n$ defined in Eq.(1). 
The event average value $\overline {{\tilde v}_n}$ of these flow coefficients 
${\tilde v}_n$, is obtained by averaging ${\tilde v}_n$ over 
a large number of events (of similar type, e.g. centrality etc.). 
$\overline {{\tilde v}_n}$ is determined by the following expression

\begin{equation}
 \overline {{\tilde v}_n} =  \frac{1}{2\pi} \int_{-\pi}^\pi (\frac{1}{2\pi} 
\int_{-\pi}^\pi  \rho_\psi(\phi) e^{-in\phi} d\phi ) d\psi  
\end{equation}

 Integration over $\psi$ reflects the random variation of the orientation 
of the event plane in the lab fixed frame from  event to event.
By changing the order of integration for $d\phi$ and $d\psi$, and using
Eq.(2), we immediately see that the $\overline {{\tilde v}_n} = 0$ for 
all $n > 0$ as mentioned above. 

  We next calculate the event averaged values of $|{\tilde v}_n(\psi)|^2$
as

\begin{equation}
 ({\tilde v}_n^{rms})^2 = \overline {{\tilde v}_n^2} =  \frac{1}{2\pi} \int_{-\pi}^\pi d\psi 
{\tilde v_n(\psi)} {\tilde v_n^*(\psi)}
\end{equation}

 Straightforward calculation using Eqs.(2)-(5) gives

\begin{equation}
 {\tilde v}_n^{rms} = |v_n| 
\end{equation}

 This is an important result. It shows that various conventional flow 
coefficients $v_n$, defined with respect to the event plane, can be
directly calculated in the lab fixed frame in terms of ${\tilde v}_n^{rms}$.
However, one has to be careful in interpreting the above equation. The
above equation has been derived for the event averaged values. Thus the
right hand side $|v_n|^2$ also represents the event averaged value
of the conventional flow coefficients. Even if we assume that each
event is almost identical in terms of centrality selection etc. (which
it is not), still there will be random fluctuations in the fluid 
especially due to initial state fluctuations \cite{cmbhic}. If the
contributions of such fluctuations becomes dominant then the right 
hand side of the above equation will not have a neat interpretation
of relating to $|v_n|$ defined with respect to the event plane.

   Fortunately this problem does not exist for the elliptic flow $v_2$.
Below we will see, with model calculations of ${\tilde v}_2^{rms}$ 
for non-central events from HIJING, that the contribution arising 
from the elliptic shape
of the produced partons in the overlap region is very large compared 
to the contribution from the random, initial state fluctuations. This
will be seen as a prominent peak in the plot of ${\tilde v}_n^{rms}$ vs. $n$
at $n = 2$. Thus, as far as determination of the elliptic flow is concerned, 
Eq.(6) shows that calculation of ${\tilde v}_2^{rms}$ in the lab fixed
frame can be used to determine the value of $|v_2|$. Neglecting the
contributions of random fluctuations, symmetry of the elliptical shape of
the overlap region for non-central collisions will then imply that
$|v_2|$ is the same as $v_2$ conventionally defined as the coefficient of
the $cos2\phi$ term.

  For other values of $n$ the Eq.(6) has to be suitably interpreted 
with proper account of event by event random fluctuations. As we have
emphasized, values of ${\tilde v}_n^{rms}$ determined by averaging over events
should be directly used to probe the statistical properties of fluctuations
and anisotropies of the initial plasma region (as is done for CMBR).
Hydrodynamical simulations, with proper incorporation of such fluctuations
can then be used to predict the values of ${\tilde v}_n^{rms}$ whose 
comparison with data can then be used to constrain/determine various physical
inputs such as equation of state, viscosity etc.

  We now describe our method for calculating ${\tilde v}_n^{rms}$ using 
HIJING \cite{hijingp}. We first recall from ref. \cite{cmbhic} the 
estimates of initial spatial anisotropies. We start with the 
initial transverse energy density distribution for Au-Au collision at 
200 GeV/A center of mass energy from HIJING. For parton positions we use 
random locations inside the parent nucleon, (similarly, for
partons produced from the string systems, random position are used
along the line joining the two corresponding nucleons). The
transverse energy density at a given transverse position ${\vec x}$, at
proper time $\tau = \tau_{eq}$, is taken as \cite{hijing},

\begin{equation}
\epsilon_{tr}({\vec x},\tau_{eq}) = {1 \over \Delta A}
\sum_i E_{tr}^i ~F(\tau_{eq},p_{tr})~
\delta^2({\vec x} - {\vec x}^i_0 - {\vec v}^i \tau_{eq}) ~\Delta(y^i)
\end{equation}

where ${\vec x}^i_0$ denotes the initial transverse coordinates of the
$i_{th}$ parton (determined using the coordinates of the parent nucleon in
HIJING as discussed above), $E^i_{tr}$ is its transverse energy, $p_{tr}$
the transverse momentum, and ${\vec v^i}$ is its transverse velocity.
For the rapidity window we take $\Delta(y^i) = 1$ centered at $y = 0$,
\cite{hijing}. The sum over $i$ includes all partons in a small
transverse area element $\Delta A (\simeq 0.5$ fm$^2$) at position
${\vec x}$. We have included a factor $F(\tau_{eq},p_{tr}) \equiv 1/(1 + 
1/(p_{tr} \tau_{eq})^2)$ to account for the probability of formation 
of partons with zero rapidity \cite{hijing}.

We assume that the hydrodynamic description becomes applicable by 
$\tau =\tau_{eq}$, which we take to be 1 fm and calculate the 
anisotropies in the fluctuations in the spatial extent $R(\phi)$ at 
this stage, where $R(\phi)$ represents $\epsilon_{tr}$ 
weighted average of the transverse radial coordinate in the angular bin 
at azimuthal coordinate $\phi$ \cite{cmbhic}. As emphasized above, 
angle $\phi$ is taken in a lab fixed coordinate frame. 
We divide the region in 50 - 100 bins of azimuthal angle $\phi$, and 
calculate the Fourier coefficients of the anisotropies in ${\delta R}/R 
\equiv  (R(\phi) - {\bar R})/{\bar R}$ where $\bar R$ is the angular 
average of $R(\phi)$.  Note that in this way we are representing
all fluctuations essentially in terms of fluctuations in the boundary of
the initial region.  We use $F_n$ to denote Fourier coefficients for 
these spatial anisotropies, and use ${\tilde v}_n$ to denote $n_{th}$ 
Fourier coefficient of expected momentum anisotropy in ${\delta p}/p$
defined in the lab frame. Here $\delta p$ represents 
fluctuation in the momentum $p$ of the final particles from the average 
momentum, in a given azimuthal angle bin. 

 We should clarify that in the conventional analysis, $v_2$ is directly related
to the elliptical shape of the fireball, and normally one does not call
it a fluctuation. However, in our language, the elliptic shape of
the fireball itself is taken to represent a fluctuation (of large
wavelength) from isotropic case. As our analysis is carried out
in the Lab fixed frame, all shape fluctuations (including the
elliptical ones) are treated uniformly by calculating different
Fourier coefficients $F_n$.

 Again we emphasize the most important difference between the conventional
discussions of the elliptic flow and our analysis; here
one does not try to determine any special reaction plane on event-by-event
basis. A fixed coordinate system is used for calculating azimuthal
anisotropies. This is why, as discussed in Sect.II and as is seen in
the plots in ref.\cite{cmbhic}, averages of $F_n$s
(and hence of ${\tilde v}_n$s) vanish when large number of events are included
in the analysis. However, the root mean square values of $F_n$s, and hence of
${\tilde v}_n$s, are non-zero in general and contain non-trivial information. 
In fact, it is the same as the standard deviation for the distribution of
$F_n$s since the average value of $F_n$s is zero. This is what is exactly
done for the CMBR case also \cite{cmbr}. 

 One important difference from CMBR, which may be of crucial importance in 
finding any non-trivial features in the plot of ${\tilde v}_n^{rms}$ for 
RHICE, is the fact that for the CMBR case, for each $l$ mode
of the spherical harmonic, there are only $2l+1$ independent measurements
available, as there is only one CMBR sky to observe. For small
$l$ values this becomes dominant source of uncertainty, leading 
to the accuracy limited by the so called cosmic
variance \cite{cmbr}. In contrast, for RHICE, each  nucleus-nucleus
collision (with same parameters like collision energy, centrality etc.)
provides a new sample event. Also, in the absence of any special 
reflection symmetry here (which was present in the traditional elliptic 
flow analysis) all flow coefficients give non-zero contributions to 
${\tilde v}_n^{rms}$, and the $``sin"$ terms give same values as the 
$``cos"$ terms. In the plots of ${\tilde v}_n^{rms}$
below, we show the sum of these two contributions, i.e. 
square root of the sum of the squares of the $``sin"$ term and the
$``cos"$ term

\section{Results for non-central collisions}
  
  We have generated events using HIJING and we present sample results 
for Au-Au collision at 200 GeV/A center of mass energy. In all the plots,
the averages are taken over 10000 events, and  the root mean square values 
${\tilde v}_n^{rms}$ of the flow Fourier coefficients are obtained from spatial 
$F_n$s simply by using proportionality factor of 0.2.  The relation between
the Fourier coefficients of the spatial anisotropy and resulting momentum
anisotropy in our model can only be obtained using a full hydrodynamical 
simulation, with proper accounts of any surface tension, as well as factors 
such as horizon crossing etc. to properly account for the physics discussed 
here.  In the absence of such a simulation, we 
make a strong assumption here that all Fourier coefficients for momentum 
anisotropy are related to the corresponding coefficients for spatial 
anisotropy by roughly the same proportionality factor, which we take 
to be 0.2 for definiteness. (As mentioned in ref.\cite{cmbhic}, this 
choice also gave reasonably good agreement with the results for 
${\tilde v}_2^{rms}$ in the literature for (almost) central events 
\cite{cntrl}.) Although overall shape of the plot of ${\tilde v}_n^{rms}$ will 
crucially depend on the values of these proportionality constants, and 
hence may change completely from what is shown here, the qualitative 
features such as the presence of peaks may remain unaffected. (Though,
the peak positions may shift depending on these proportionality constants.) 
$F_n$s are calculated directly from the parton energy distribution at  
$t = 1$ fm for each event, within unit central rapidity window. 

Fig.1 presents plots of ${\tilde v}_n^{rms}$ for different values of impact
parameters. Solid, dashed, and dotted plots correspond to impact parameter
$b$ = 0, 5 fm, and 8 fm, respectively with the spread in $b$ taken to be
1 fm for non-zero $b$ cases. The solid plot for central collisions 
shows monotonically decreasing values of ${\tilde v}_n^{rms}$.  We emphasize
that even here the plot has non-trivial structure compared to the 
case when partons are randomly distributed inside the nuclear volume with 
uniform probability, which yields a flat plot for corresponding 
${\tilde v}_n^{rms}$.  (Though, the rise of $b = 0$ plot for small $n$ does 
not appear prominent here due to larger vertical scale of the plot.) Errors
in these plots are very small.

  Important thing to note is the prominent peak in non-central collisions
for $b$ = 5, and 8 fm at $n = 2$. The peak height is significantly larger 
for $b$ = 8 fm. Important thing to realize is that this feature (which 
within our model simply represents elliptical distribution of partons
in position space at the initial time) is present even when the 
${\tilde v}_n^{rms}$ have been calculated in a lab fixed reference frame. 
Average values of ${\tilde v}_n$
for all values of impact parameters, calculated for large number of events,
continue to be zero as shown in ref. \cite{cmbhic} for the central 
collision case. This is expected from rotational symmetry in lab fixed frame.
Larger overall values of ${\tilde v}_n^{rms}$ for larger values of $b$ are 
simply due to smaller number of initial partons leading to larger fluctuations.
We mention that the solid plot in Fig.1 is given exactly for $b = 0$
to show the difference between zero and non-zero $b$ cases. 
However, with experimental data, $b$ = 0 will inevitably 
include events with small but non-zero values of $b$. Thus even for 
$b = 0$ case one should expect to observe a (small) peak at $n = 2$.

\begin{figure}
\vskip -1.5in
\epsfig{file=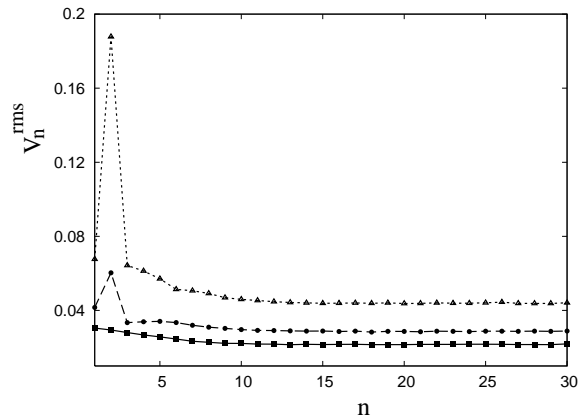,height=130mm}
\vskip -1.5in
\caption{These plots show smooth joining of the values of ${\tilde v}_n^{rms}$ 
calculated from Fourier coefficients $F_n$s of the initial spatial 
anisotropy as discussed in text. Plots are obtained using 10000 events 
from HIJING. Solid, dashed, and dotted plots correspond to impact parameter
$b$ = 0, 5 fm, and 8 fm, respectively. Note the prominent peak at $n = 2$ for
non-zero b cases.}
\label{fig:fig1}
\end{figure}

We have also calculated the root mean square values of the Fourier 
coefficients of the momentum anisotropy directly using the momenta of
final particles from HIJING. These plots are all similar to each other 
for all values of the impact parameter $b$ and hence similar to the central 
case $b = 0$ (apart from larger statistical fluctuations for larger values 
of $b$ due to smaller number of partons as discussed above for Fig.1.) 
This is expected because, despite initial spatial anisotropy of parton 
production for non-central collisions, momentum distribution remains 
isotropic in the absence of any hydrodynamic evolution. We do not show 
these momentum plots here because such plots have been given in 
ref.\cite{cmbhic} for the central collision case.

 Fig.1 shows the strength of this technique of plotting root-mean-square
values of flow coefficients in lab fixed coordinate frame as compared 
to the traditional analysis of elliptic flow. This technique is  simple to 
implement (apart from the issue of non-flow correlations 
\cite{nonflow,accpt1}, which one has to resolve in any case), allowing larger 
statistics to be generated. What one needs is predictions from hydrodynamics 
about ${\tilde v}_n^{rms}$ calculated in lab fixed frame and then compare with
data analyzed similarly. As the plots in Fig.1 show, the values of
${\tilde v}_n^{rms}$ even for very large values of $n$ continue to be 
significantly non-zero and with the possibility of sufficient statistics, 
one may be able to determine any important features in these plots.

  In \cite{cmbhic} we had argued that certain important aspects of
inflationary density fluctuations, such as suppression of superhorizon
modes and acoustic oscillations may also be present in these plots  of 
${\tilde v}_n^{rms}$ in RHICE. Basic ideas can be stated as follows. Acoustic 
peaks in CMBR primarily result from the coherence and acoustic oscillations
of the inflationary density fluctuations with coherence originating
from the superhorizon nature of fluctuations, essentially freezing
the fluctuations on superhorizon scales. This should be reasonably 
true for RHICE as the transverse velocity to begin with
is expected to be zero and becomes non-zero only due to pressure
gradients (which does not become fully effective on superhorizon scales).
The oscillatory behavior for the fluctuations in the universe results
from attractive forces of gravity and counter balancing forces from
radiation pressure (with the coupling of baryons to the radiation). 
The oscillatory behavior of the fluctuations in RHICE is expected due to
flow development with non-zero pressure  gradients, at least 
for sufficiently small wavelength modes.
These arguments suggest that some features like acoustic peaks may be 
present in the plots of ${\tilde v}_n^{rms}$ for
RHICE also. Further, modes with wavelengths larger than the acoustic horizon
size $H^{fr}_s$ at freezeout should be naturally suppressed as pressure 
gradients do not become fully effective on those scales to generate full
flow anisotropy before freezeout occurs. This leads to the following
suppression factor 
 
\begin{equation}
({\tilde v}_n)_{observed} = {2H^{fr}_s \over \lambda} ({\tilde v}_n)_{max}
\end{equation}

where $\lambda \sim 2\pi {\bar R}^{fr}/n, ~~~(n \ge 1)$, is the measure of 
wavelength of the anisotropy corresponding to the $n_{th}$ Fourier coefficient.
Here $\bar R^{fr}$ represents the transverse radius at the freezeout time
$\tau_{fr}$.
Using the rough estimate of the rate of change of the transverse velocity
to be about 0.1 fm$^{-1}$ at the early stages at these energies \cite{vtr},
we can estimate $\bar R^{fr} \simeq {\bar R} + 0.05 (\tau_{fr} -
\tau_{eq})^2 = {\bar R} (1 + 0.05 {\bar R}/c_s^2)$. Here ${\bar R} \equiv
{\bar R}(\tau_{eq}) = c_s (\tau_{fr} - \tau_{eq})$.
The largest wavelength $\lambda_{max}$ of spatial anisotropy which will
have chance  to develop to its maximum hydrodynamic value is, therefore,
$\lambda_{max} \simeq 2 H^{fr}_s = 2 c_s (\tau_{fr} - \tau_{eq})
= 2 {\bar R}(\tau_{eq})$. This gives us the corresponding minimum value
$n_{min}$ of $n$ below which flow coefficients should show suppression
due to being superhorizon,

\begin{equation}
n_{min} = \pi (1 + {0.05 {\bar R}(\tau_{eq}) \over c_s^2} )
\end{equation}

  The possibilities of the superhorizon suppression  and acoustic 
oscillations lead to new peak structures in the plot of ${\tilde v}_n^{rms}$
as was shown in \cite{cmbhic} for central collisions. It is therefore
important to know whether the peak in ${\tilde v}_n^{rms}$ due to elliptic 
flow for non-central collisions (as in Fig.1) survives if these new features 
are also present, especially with the suppression of superhorizon modes at
smaller $n$. For this purpose, we show in Fig.2, plots (for different
values of impact parameter $b$ as in Fig.1), when superhorizon suppression is
included and when acoustic oscillations are also included in modeling the
values of ${\tilde v}_n^{rms}$. It is clearly seen that the peak at $n = 2$
corresponding to the elliptic flow is present in all the plots
and remains most prominent for $b = 8$ fm.

 Here we mention
that recently Sorensen has proposed \cite{srnsn} a novel explanation of
azimuthal correlations observed at the Relativistic Heavy-ion Collider 
(RHIC) at BNL in terms of the suppression of the superhorizon fluctuations
discussed in \cite{cmbhic}. One of the plots (Fig.4) in \cite{srnsn}
(where suitable subtraction of $v_2$ has been made) shows possible 
suppression of values of the variance for small $n$, as predicted in
\cite{cmbhic}. 

\begin{figure}
\vskip -0.4in
\epsfig{file=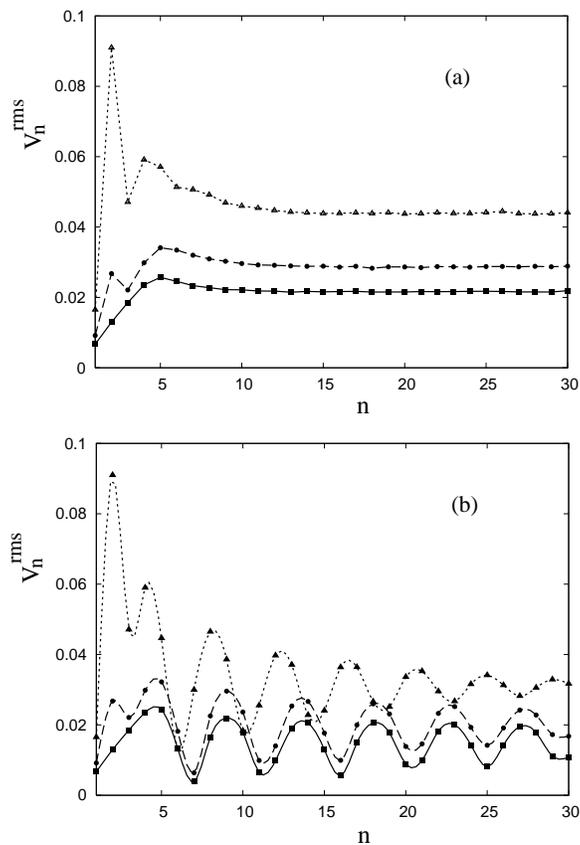,height=130mm}
\vskip -0.3in
\caption{These plots are obtained from the values of ${\tilde v}_n^{rms}$ 
of the plots in Fig.1 by including (a) the suppression factor given in Eq.(8)
for superhorizon modes, and (b) including acoustic oscillations as well. 
The plots here only show the modeling of the type 
of suppression  factor and oscillations discussed in the text, which 
are superimposed on the plots in Fig.1.}
\label{fig:fig2}
\end{figure}

 We mention that we calculate ${\tilde v}_n^{rms}$ by direct calculation of 
variances of the distributions of ${\tilde v}_n$ (actually $F_n$s) in the 
laboratory fixed frame. Equivalently one can calculate ${\tilde v}_n^{rms}$ 
by calculating the two-particle azimuthal correlation functions, as is 
done for the elliptic flow calculations. To extract elliptic flow 
coefficient from two-particle azimuthal correlations one has to separate 
non-flow contributions \cite{nonflow,accpt1}. Two particle azimuthal 
correlations, which are experimentally measured, 
contain contributions from non-flow effects such 
as jets, resonance decays, HBT correlations, final state interactions etc. 
Various methods have been discussed to separate out the non-flow
contributions to the azimuthal correlations \cite{nonflow,accpt1}. 
Our estimates of ${\tilde v}_n^{rms}$ will also contain such non-flow 
contributions which have to be properly accounted for.

\section{$p_T$ dependence}

  In the conventional analysis of elliptic flow, $v_2$ has a non-trivial
dependence on $p_T$ of hadrons \cite{v2pt}. $v_2$ increases with $p_T$,
saturates, and then decreases. Initial increase of $v_2$ with $p_T$ is easily
determined \cite{sv,oltr,oltr1}. Large $p_T$ behavior of flow is governed by 
viscous effects, parton energy  loss, as well as by parton coalescence 
\cite{v2pt}. The same physics will be applicable in our model as well. 
For simplicity, we will follow the approach in ref. \cite{oltr}
where the linear increase of the elliptic flow $v_2$ with $p_T$ 
is determined to be

\begin{equation}
v_2 = {\alpha \over T}(p_T - v m_T)
\end{equation} 

where $\alpha$ characterizes the magnitude of elliptic flow, $v$ is the
flow velocity (see below) and $m_T = \sqrt{p_T^2 + m^2}$ 
is the transverse mass. We mention that this linear dependence of 
$v_2$ on $p_T$ is expected for the intermediate range of $p_T$. The 
behavior for small $p_T$ may be more non-trivial \cite{sv}. However,
recall from Eq.(6) (and discussion in Sect.II) that in the context 
of our model, ${\tilde v}_2^{rms}$ is simply given by the conventional 
$v_2$, hence whatever be the dependence of $v_2$ on $p_T$, 
${\tilde v}_2^{rms}$ should show the same dependence. Still, it is useful
to directly check the expected $p_T$ dependence of ${\tilde v}_2^{rms}$ in
our model by adopting the analysis of $v_2$ in ref. \cite{oltr} for our case.
This is for two reasons. First, the relevance of random fluctuations needs
to be assessed at different stages of the calculations to see the effects
of event averaging and to see  whether similar statements can be made
about other flow coefficients. (For example, see the discussion following 
Eq.(12) below about the form of the fluid 4-velocity.) Secondly, the
analysis for $v_2$ in ref. \cite{oltr} is carried out to linear order
in $\alpha$. Since we calculate the root mean square values, we need to
keep terms of order $\alpha^2$. Though, as we will see below, the $\alpha^2$
term will turn out to be only relevant for $v_4$ and will not affect the 
analysis of ${\tilde v}_2^{rms}$.

 We start by noting that the expression for $v_2$ in Eq.(10) corresponds to 
the conventional 
analysis with the determination of event plane. In our approach, the 
analysis is carried out in the lab fixed frame. For each non-central event 
the magnitude of the flow anisotropy will still be characterized by (say) 
the above equation. However, the maximum flow direction will now vary 
randomly from one event to another because the event plane is
left undetermined.  Thus, it looks reasonable to expect that 
${\tilde v}_2^{rms}$ should show the same dependence on $p_T$ as $v_2$ in 
Eq.(10), as is indeed shown by Eq.(6). We will  check this directly by 
adopting the analysis of $v_2$ in ref. \cite{oltr} for our case for the 
estimation of ${\tilde v}_2^{rms}$ in the lab fixed frame (and keeping 
terms of order $\alpha^2$). Let us start with
the transverse momentum distribution of emitted particles \cite{oltr}

\begin{equation}
{dN \over p_T dp_T dp_z d\phi} \equiv \rho(\phi) \propto 
exp({-m_T u^0(\phi) + p_T u(\phi) \over T})
\end{equation} 

Here, $u(\phi)$ is the space component of the fluid flow 4-velocity
$u_\mu$ in $\phi$ direction, and $u^0(\phi)$ is the time component.

 For non-central collisions, the anisotropic flow can be parameterized as

\begin{equation}
u(\phi) = u + 2\alpha cos 2(\phi - \psi)
\end{equation} 

 Here $u$ is the angular average of maximum fluid
4-velocity, and $\alpha > 0$ characterizes the magnitude of elliptic flow.
Note, that we have introduced here the angle $\psi$ characterizing the 
orientation of the event plane in the lab frame so that the angle $\phi$ is
measured in the lab frame. Also, it is important to realize that here
we are only characterizing elliptic flow anisotropy of the fluid, without
worrying about the presence of general fluctuations. The justification for 
this is that, as can be seen from Fig.1, for non-central collisions
in our method it is only ${\tilde v}_2^{rms}$ which has a significantly larger 
value than the values of ${\tilde v}_n^{rms}$ for other $n$. Thus, the 
presence of general fluctuations (as accounted for in Fig.1) will play an 
important role in determination of these other values of ${\tilde v}_n^{rms}$ 
even for non-central collisions. In contrast, for ${\tilde v}_2^{rms}$ the 
elliptic anisotropy resulting from non-central collision seems to play 
dominant role over the random fluctuations. Hence the parametrization of 
Eq.(12) may capture important aspects of the dependence of 
${\tilde v}_2^{rms}$ on $p_T$ etc.    

  We are interested in calculating the root mean square value (instead of $v_2$
as in ref. \cite{oltr}). Thus, we expand $u^0(\phi) = \sqrt{u(\phi)^2 + 1}$
to second order in $\alpha$, and get (with $v = u/u^0$), 

\begin{equation}
u^0(\phi) = u^0 + 2\alpha v cos2(\phi - \psi) + {2 \alpha^2 \over (u^0)^3}
cos^22(\phi - \psi)
\end{equation} 

Using this equation, and Eq.(11), we get particle transverse momentum
distribution as 

\begin{equation}
\rho_\psi(\phi) = A[1 + 2 \alpha cos2(\phi-\psi) {(p_T - v m_T) \over T}
+ \alpha^2 (1 + cos4(\phi - \psi)) ({(p_T - v m_T)^2 \over T^2} 
- {m_T \over T (u^0)^3} ) ]
\end{equation} 
 
where $A$ is independent of $\phi, \psi$. We see that this equation is
in the same form as Eq.(2). It is important to note here that the 
$\alpha^2$ term here only comes for $v_4$ (requiring higher powers of 
$\alpha$ for the calculation of ${\tilde v}_4^{rms}$) and $v_2$ remains 
linear in $\alpha$. 

Following the discussion in
Sect.II following Eq.(2), and using Eq.(6), we immediately see that

\begin{equation}
{\tilde v}_2^{rms} =  {\alpha \over T} (p_T - v m_T)
\end{equation} 

Which is the same dependence on $p_T$ as in Eq.(10) for the conventional 
analysis of $v_2$ using event plane determination.

 We again emphasize that in our picture, the randomness in the orientation 
of the event plane w.r.t. the lab fixed frame is easy to model by introducing 
angle $\psi$. The other intrinsic sources of fluctuations will also 
contribute to ${\tilde v}_2^{rms}$ in our method, but as  shown in Fig.1, 
their contributions appear to remain small (as seen by the plot for $b$ = 0). 
(Note again, overall larger scale of plots of ${\tilde v}_n^{rms}$ for
larger values of $b$ is primarily due to smaller number of particles leading 
to larger statistical fluctuations.) As we had discussed earlier, 
${\tilde v}_n^{rms}$ for $n \ne 2$ may be dominated by these intrinsic 
sources of fluctuations. If the contribution of such fluctuations remains 
sub dominant, then our results show that direct determination of 
$v_2$, in particular its $p_T$ dependence (\cite{sv}) is possible 
by measuring ${\tilde v}_2^{rms}$ in lab fixed frame completely avoiding 
the determination of event plane.

\section{detector acceptance}

  One of the main advantages of using the techniques proposed here is
that it makes it much easier to extract the information about flow
without any need for event plane determination. Thus large amount of data 
can be analyzed improving accuracy. It then becomes important to address
the issue of being able to incorporate data with varying azimuthal
acceptance, as has been emphasized for the conventional flow analysis. 
Several methods have been proposed for accounting for azimuthal
dependence of detector acceptance in the conventional flow analysis 
\cite{accpt1,accpt2}. These can be suitably adopted to the method proposed
here. As mentioned above, due to fixed lab frame being
used for analysis, the average values of all flow coefficients will
be zero here due to azimuthal symmetry. Clearly, this will  no longer
be true in the presence of azimuthal anisotropy in the detector 
acceptance. For example, let $A(\phi)$ be the acceptance function
characterizing the probability that a hadron is detected at azimuthal 
angle $\phi$, with $A(\phi)$ being normalized as \cite{accpt1,accpt2}

\begin{equation}
\int^{\pi}_{-\pi} A(\phi) d\phi = 2\pi
\end{equation} 

 In the lab fixed frame, the values of ${\tilde v}_n$ in our method may be 
calculated from Eq.(3) (Sect.II). Note that
for anisotropic detector acceptance there is an implicit factor
of $A(\phi)$ in the integrand in Eq.(3). One can then directly 
calculate the event average value $\overline {{\tilde v}_n}$ of these flow  
coefficients ${\tilde v}_n$, by averaging ${\tilde v}_n$ over a large 
number of events (of similar type, e.g. centrality etc.). The actual value
of $\overline {{\tilde v}_n}$, for non-central collisions, is determined by
the following expression

\begin{equation}
 \overline {{\tilde v}_n} =  \frac{1}{2\pi} \int_{-\pi}^\pi (\frac{1}{2\pi} 
\int_{-\pi}^\pi A(\phi) 
\rho_\psi(\phi) e^{-in\phi} d\phi ) d\psi  
\end{equation}

 Here, $\psi$ denotes the orientation of the event plane in the lab fixed
frame (see, Eq.(4) in Sect.II), which varies from event to event. We have 
made here explicit the acceptance function $A(\phi)$ in the above 
expression. (For simplicity we assume that $A(\phi)$ does not depend on 
rapidity, $p_T$ etc.). For ideal detector with uniform acceptance 
$A(\phi) = 1$ and by changing the order of $d\phi$ and $d\psi$ integrations 
one can easily see that $\overline {{\tilde v}_n} = 0$ as discussed in Sect.II
(following Eq.(4)). However, for anisotropic acceptance, with $A(\phi) 
\ne 1$, $\overline {{\tilde v}_n}$ need not be zero. In fact, apart from a 
normalization (relating to angular average of $\rho$), $\overline 
{{\tilde v}_n}$
gives the Fourier coefficients for the series expansion of the function
$A(\phi)$, thus allowing us to determine the acceptance function 
$A(\phi)$ directly from experimental data \cite{accpt1,accpt2}.

For the simple case when  $A(\phi)$ is non-zero for every $\phi$, one can 
directly use $A(\phi)$, determined from $\overline {{\tilde v}_n}$ as explained
above, to compensate for the effect of anisotropic detector acceptance 
by including additional factor of $1/A(\phi)$ in the integrand of 
Eq.(3) for calculation of ${\tilde v}_n$. 

\begin{equation}
 {\tilde v}_n =  \frac{1}{2\pi} \int_{-\pi}^\pi 
{\rho_\psi(\phi) \over A(\phi)} e^{-in\phi} d\phi 
\end{equation}

With this modification, Eq.(17) shows that the event average value 
$\overline {{\tilde v}_n}$ will again be zero, just as
for the case of isotropic detector acceptance. The above Eq.(18) can
then be used to determine the values of ${\tilde v}_n^{rms}$ as discussed
in earlier sections. 

 However, for more general case, e.g. for incomplete detector coverage, 
where $A(\phi) = 0$ for a range of $\phi$ the above simple method
does not help. For such cases one can adopt the techniques discussed in
\cite{accpt1,accpt2} for the present case. Though, due to the important
role played by the event plane determination, these techniques need to
be suitably adopted for our case where a fixed lab frame is used for
the analysis. In fact, it is more helpful to directly adopt the techniques
used for CMBR case where one uses a fixed lab frame (say, galactic
coordinates) for writing down temperature anisotropies in the sky in
terms of spherical harmonics. For CMBR analysis also one needs to compensate 
for the effects of partial coverage of the sky. This happens either due to
limited coverage of the sky by the detector, or due to galactic
foreground as for WMAP \cite{cmbrprtl}. Adopting that approach for
our case, we start with the experimentally determined flow coefficients
${\rm v}_n$ for each event (which, following   \cite{cmbrprtl} we 
call {\it pseudo} flow coefficients).

\begin{equation}
 {\rm v_n} =  \frac{1}{2\pi} \int_{-\pi}^\pi e^{-in\phi} A(\phi)
\rho_\psi(\phi) d\phi
\end{equation}

 where, now, $A(\phi)$ may be zero for a range of $\phi$ values. 
Using Fourier coefficients $a_m$ for the acceptance function
$A(\phi)$, we get,

\begin{equation}
 {\rm v_n} =  \frac{1}{2\pi} \int_{-\pi}^\pi \sum_m a_m e^{-i(n-m)\phi} 
\rho_\psi(\phi) d\phi ~~ = \sum_l a_{(n-l)}~ {\tilde v}_l 
\end{equation}

 where ${\tilde v}_l$ is the true flow coefficient (defined with
respect to the lab fixed frame). The above equation
shows the mode-mode coupling of flow coefficients resulting
from incomplete detector coverage \cite{cmbrprtl,accpt1,accpt2}.
For an approximate determination of ${\tilde v}_l$ we can use appropriate
truncation of the above matrix equation to a finite set of
linear equations and solve for ${\tilde v}_l$. This analysis can be done
for each event, and with the values of ${\tilde v}_l$ thus determined one
can directly get the estimate of the true variance ${\tilde v}_n^{rms}$
for a set of events.

\section{Conclusions}

 In summary we emphasize the important lessons from CMBR
analysis techniques. We have argued that important information
about initial anisotropies of the system and their evolution in relativistic
heavy-ion collisions can be obtained by plotting the root mean square values 
of the Fourier coefficients ${\tilde v}_n^{rms}$ of the anisotropies in the 
fluctuations $\delta p/p$ of the particle momenta, calculated in a fixed
laboratory frame, starting from $n = 1$ upto large values of $n \simeq 30$. 
(which corresponds to  $\lambda \sim 1$ fm). 
Note that $n = 30$ almost corresponds to wavelength of fluctuation $\lambda$
at the surface of the region, at $\tau_{fr}$,
being of order 1 fm.  Fluctuations with wavelengths smaller than 1 fm 
presumably cannot be treated within hydrodynamical
framework, so we restrict attention within this range of $n$.
One will expect that beyond a critical value of $n$ the nature of the
curve should change in some qualitative manner indicating breakdown
of underlying hydrodynamical description for smaller modes. The wavelength
corresponding to that critical value of $n$ will determine the smallest
scale below which hydrodynamical description is not valid. The plot of 
${\tilde v}_n^{rms}$ for $n$ larger than this critical value will probe
fluctuations in parton density at even smaller length scales and may
provide a bridge with the perturbative regime.

 For non-central collisions our technique provides a direct method to
probe elliptic flow by determining ${\tilde v}_2^{rms}$ in lab fixed frame.
If random fluctuations do not dominate then our results show that
${\tilde v}_n^{rms}$ can directly probe $|v_n|$ for $n \ne 2$ also. This
may be important for collisions of deformed nuclei. It is important
to appreciate that a plot of ${\tilde v}_n^{rms}$ tells us about the 
statistical nature of fluctuations and anisotropies of the initial plasma 
region (just as in the CMBR case), hence it will always have valuable 
information irrespective of its shape (e.g. a flat curve). 
Hydrodynamical simulations, with proper incorporation of such fluctuations
can  be used to predict directly the values of ${\tilde v}_n^{rms}$ 
whose comparison with data will constrain/determine various physical
inputs such as equation of state, viscosity etc.  Further, in
analogy with CMBR analysis, one can determine higher moments of momentum 
anisotropies (again in lab fixed frame, as we have discussed above)
probing detailed nature of initial fluctuations, e.g. non-Gaussianity etc.
We hope to present a more detailed analysis of such issues in a future work.
We also intend to explore various aspects of {\it horizon entering}
of density fluctuations which can be probed under laboratory conditions
using RHICE. This may help in bringing at least some aspects of inflationary
physics (e.g. causal aspects, excluding those features which relate
directly to gravity) of the universe under some experimental control.

\section*{Acknowledgments}

  We are very grateful to Raju Venugopalan, Rajeev Bhalerao, and Sergei
Voloshin for very useful comments and suggestions. We also thank 
Sanatan Digal, Abhishek Atreya and Anjishnu Sarkar for useful comments.

%%%%%%%%%%%%%%%%%%% 


\begin{thebibliography}{99}

\bibitem{cmbhic}  A. P. Mishra, R. K. Mohapatra, P. S. Saumia, and 
A. M. Srivastava, Phys. Rev. {\bf C 77}, 064902 (2008).

\bibitem{hijing} M. Gyulassy, D. H. Rischke, and B. Zhang,
Nucl. Phys. {\bf A  613}, 397 (1997).

\bibitem{flow0} J.-Y. Ollitrault, Phys. Rev. {\bf D 46}, 229 (1992);
S. Voloshin and Y. Zhang, Z. Physik {\bf C70}, 665 (1996);
S.A. Volosin, A.M. Poskanzer, and R. Snellings, arxiv:0809.2949.

\bibitem{cntrl} B. M. Tavares, H. J. Drescher, and T. Kodama,
Braz. J. Phys. {\bf 37}, 41 (2007); P. Sorensen (for the STAR 
collaboration), J. Phys. {\bf G 34}, S897 (2007); S. Manly et al. 
(for PHOBOS collaboration), nucl-ex/0702029; H. Sorge, 
Phys. Rev. Lett. {\bf 82}, 2048 (1999).

\bibitem{cmbr} S. Dodelson, "Modern Cosmology", (Academic
Press, California, 2003).

\bibitem{hijingp} X.N. Wang and M. Gyulassy, Phys. Rev. {\bf D 44},
3501 (1991); Comput. Phys. Commun. {\bf 83}, 307 (1994).

\bibitem{nonflow} N. Borghini and J.-Y. Ollitrault, Phys. Rev. 
{\bf C 70}, 064905 (2004); X. Dong, S. Esumi, P. Sorensen, N.Xu,
and Z. Xu, Phys. Lett. {\bf B 597}, 328 (2004)

\bibitem{accpt1} N. Borghini, P. M. Dinh, and J.-Y. Ollitrault, 
Phys. Rev. {\bf C 64}, 054901 (2001). 

\bibitem{vtr} P.F. Kolb, J. Sollfrank, and U. Heinz,
Phys. Rev. {\bf C 62}, 054909 (2000).

\bibitem{srnsn} P. Sorensen, Proc. 24th Winter Workshop on
Nuclear Dynamics, 2008, arXiv:0808.0503.

\bibitem{v2pt} M. Gyulassy, I. Vitev, X.N. Wang,
Phys.Rev.Lett. {\bf 86}, 2537 (2001); Y. Bai (for STAR collaboration),
J.Phys. {\bf G34}, S903 (2007); B.I. Abelev et al. (for STAR
Collaboration), Phys.Rev. {\bf C77}, 054901 (2008).

\bibitem{oltr} J.-Y. Ollitrault, Eur. J. Phys. {\bf 29}, 275 (2008). 

\bibitem{oltr1} N. Borghini, and J.-Y. Ollitrault, Phys. Lett.
{\bf B 642}, 227 (2006). 

\bibitem{sv} P. Huovinen, P.F. Kolb, U. Heinz, P.V. Ruuskanen,
and S.A. Voloshin, Phsy. Lett. {\bf 503}, 58 (2001).

\bibitem{accpt2} I. Selyuzhenkov and S. Voloshin, Phys. Rev.
{\bf C 77}, 034904 (2008); R. S. Bhalerao, N. Borghini, and J.-Y. 
Ollitrault, Nucl. Phys. {\bf A 727}, 373 (2003). 

\bibitem{cmbrprtl} E. Hivon, et al. Astrophys. J. {\bf 567},2 (2002). 

\end{thebibliography}
\end{document}